\documentclass[letterpaper]{article} 
\usepackage{aaai2026}  
\usepackage{times}  
\usepackage{helvet}  
\usepackage{courier}  
\usepackage[hyphens]{url}  
\usepackage{graphicx} 
\usepackage{natbib}  
\usepackage{caption} 
\usepackage{algorithm}
\usepackage{algorithmic}

\usepackage{newfloat}
\usepackage{listings}
\DeclareCaptionStyle{ruled}{labelfont=normalfont,labelsep=colon,strut=off} 
\floatstyle{ruled}
\newfloat{listing}{tb}{lst}{}
\floatname{listing}{Listing}
\usepackage{makecell}
\usepackage{multirow}
\usepackage{array}
\usepackage{booktabs}
\usepackage{longtable}
\usepackage{pdflscape}
\usepackage{graphicx} 
\usepackage{calc}

\usepackage{pifont}
\newcommand{\cmark}{\ding{51}}      
\newcommand{\xmark}{\ding{55}}      
\newcommand{\ocircle}{\raisebox{0.1ex}{\large$\circ$}}

\title{HPSU: A Benchmark for Human-Level Perception \\in Real-World Spoken Speech Understanding}

\author{
    Chen Li\textsuperscript{\rm 1}\equalcontrib,
    Peiji Yang\textsuperscript{\rm 2}\equalcontrib,
    Yicheng Zhong\textsuperscript{\rm 2}\equalcontrib,
    Jianxing Yu\textsuperscript{\rm 1,\rm 3}\thanks{Corresponding author: yujx26@mail.sysu.edu.cn},
    Zhisheng Wang\textsuperscript{\rm 2}, \\
    Zihao Gou\textsuperscript{\rm 1},
    Wenqing Chen\textsuperscript{\rm 4},
    Jian Yin\textsuperscript{\rm 1}
}

\affiliations{
    \textsuperscript{\rm 1}School of Artificial Intelligence, Sun Yat-sen University, Zhuhai, 519082, China\\
    \textsuperscript{\rm 2}Tencent, Shenzhen, 518000, China\\
    \textsuperscript{\rm 3}Key Laboratory of Sustainable Tourism Smart Assessment Technology, Ministry of Culture and Tourism, 519082, China\\
    \textsuperscript{\rm 4}School of Software Engineering, Sun Yat-sen University, Zhuhai, 519082, China\\
    \{lich528, yujx26, gouzh, chenwq95, issjyin\}@mail.sysu.edu.cn,
    \{peijiyang, ajaxzhong, plorywang\}@tencent.com
}

\begin{document}

\maketitle

\begin{abstract}
Recent advances in \textit{Speech Large Language Models} (\textit{Speech LLMs}) have led to great progress in speech understanding tasks such as \textit{Automatic Speech Recognition} (\textit{ASR}) and \textit{Speech Emotion Recognition} (\textit{SER}). However, whether these models can achieve human-level auditory perception, particularly in terms of their ability to comprehend latent intentions and implicit emotions in real-world spoken language, remains underexplored. To this end, we introduce the \textbf{H}uman-level \textbf{P}erception in \textbf{S}poken \textbf{S}peech \textbf{U}nderstanding (\textit{HPSU}), a new benchmark for fully evaluating the human-level perceptual and understanding capabilities of \textit{Speech LLMs}. \textit{HPSU} comprises over 20,000 expert-validated spoken language understanding samples in English and Chinese. It establishes a comprehensive evaluation framework by encompassing a spectrum of tasks, ranging from basic speaker attribute recognition to complex inference of latent intentions and implicit emotions. To address the issues of data scarcity and high cost of manual annotation in real-world scenarios, we developed a semi-automatic annotation process. This process fuses audio, textual, and visual information to enable precise speech understanding and labeling, thus enhancing both annotation efficiency and quality. We systematically evaluate various open-source and proprietary \textit{Speech LLMs}. The results demonstrate that even top-performing models still fall considerably short of human capabilities in understanding genuine spoken interactions. Consequently, \textit{HPSU} will be useful for guiding the development of \textit{Speech LLMs} toward human-level perception and cognition.
\end{abstract}

\begin{links}
    \link{Code}{https://github.com/Ichen12/HPSU-Benchmark}
\end{links}

\section{Introduction}

Speech processing is a hot research topic. This technology has greatly advanced the development of human-computer interaction, moving beyond basic tasks like \textit{ASR} toward genuine comprehension~\cite{ji2024wavchat}. Humans' expectation for this technology's performance has also increased. They require it to be able to perceive, comprehend, and infer the fine-grained, implicit, and multidimensional nuances inherent in human speech~\cite{arora2025landscape}.

Human auditory comprehension is an intricate process that encompasses not only transcription but also a holistic grasp of speaker attributes, latent intentions, and complex emotional states.~\cite{hellbernd2016prosody} While current models have shown good performance on isolated tasks, a significant gap persists in evaluation methodologies. The preceding benchmarks have mostly focused on coarse-grained tasks or textual reasoning, neglecting the integrated perceptual abilities that define human-level understanding. Furthermore, reliance on non-interactive data and frequent confinement to a single language preclude robust assessment of model capabilities in authentic, multilingual communicative contexts~\cite{yu2025eliciting}.

To address this problem, we introduce the \textbf{H}uman-level \textbf{P}erception in \textbf{S}poken Speech \textbf{U}nderstanding (\textit{HPSU}) benchmark. \textit{HPSU} is a distinctive evaluation framework comprising over 20,000 expert-validated instances in both Chinese and English. It is carefully designed to fully assess the deep perceptual and cognitive abilities of \textit{Speech LLMs}. Its construction was enabled by an effective annotation pipeline that emulates multimodal human cognition. This pipeline fuses audio, textual, and visual information to achieve high-fidelity labeling of subtle communicative signals at scale. The benchmark's design features a hierarchical taxonomy of 16 tasks that probe advanced challenges largely absent from prior work, such as tracking emotional dynamics and inferring conversational subtext. Methodologically, \textit{HPSU} incorporates key innovations, including a sophisticated distractor-generation mechanism for subjective tasks and a dedicated adversarial protocol to assess model robustness against misleading information.

Our primary contributions are threefold:
\begin{itemize}
    \item We propose \textit{HPSU}, a large-scale benchmark for evaluating in-depth speech understanding, encompassing over 20,000 instances across 16 diverse tasks.
    \item We introduce a multimodal annotation pipeline that greatly reduced the annotation cost, enabling the efficient and high-quality construction of our benchmark, and release the \textit{HPSC} dataset, which contains 50,000 high-quality speech-description pairs in English and Chinese.
    \item We conduct extensive evaluation on 13 leading models based on \textit{HPSU}. The results reveal a substantial deficit compared to human performance, particularly in tasks of latent semantic perception and complex emotion reasoning, thereby charting a clear path for future research.
\end{itemize}

\begin{figure*}[!ht]
\centering
\includegraphics[width=0.85\textwidth,keepaspectratio]{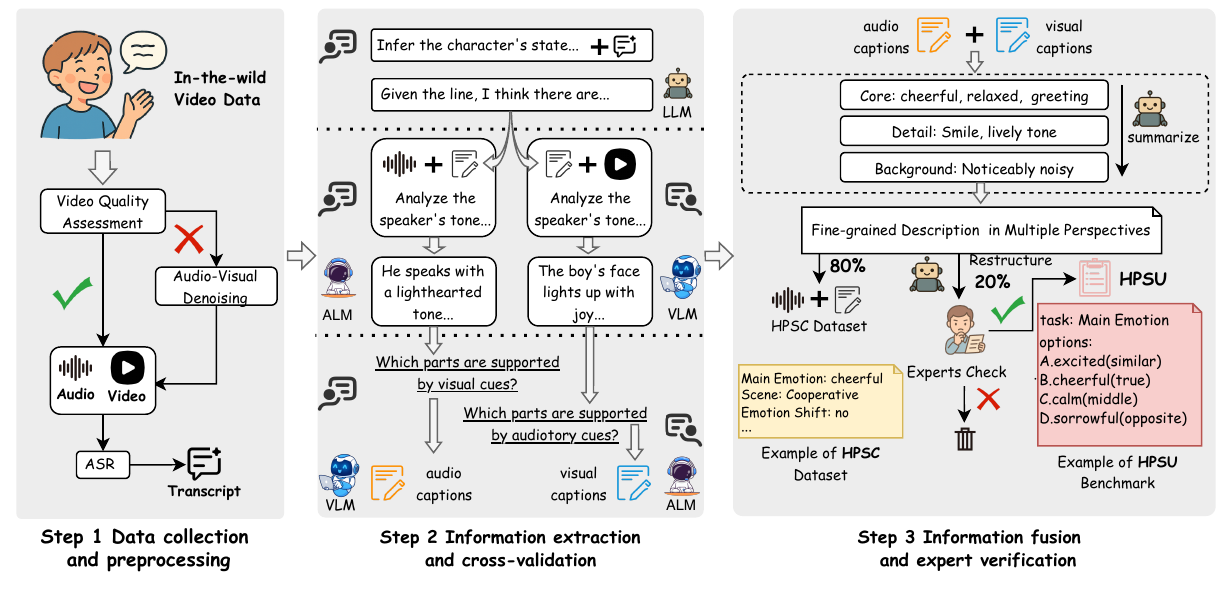}
\caption{Overview of the data construction and annotation pipeline for \textit{HPSC} dataset and \textit{HPSU} benchmark.}
\label{fig:pipeline}
\end{figure*}

\section{Method}

\subsection{Construction and Annotation Pipeline}

The construction of \textit{HPSU} is grounded in a large-scale corpus of authentic spoken data. The data was sourced from thousands of hours of public audio-visual datasets rich in semantic and stylistic detail. To overcome the efficiency and quality bottlenecks of traditional methods, we designed a semi-automatic annotation pipeline. As shown in Figure \ref{fig:pipeline}, there are three stages in our pipeline:

\textbf{Step 1: Data collection and preprocessing.} We first select high-quality clips from multiple real-world video corpora, including \textit{CelebV-HQ}~\cite{zhu2022celebvhq}, \textit{MAFW}~\cite{liu2022mafw}, \textit{MELD}~\cite{poria2019meld}, and \textit{MER}~\cite{lian2025mer2025affectivecomputing}. These data are mainly sourced from open-domain scenes such as movie clips and social media videos, which possess advantages such as diverse scenarios, rich emotional expressions, and varied, natural speaking styles. These properties provide a solid foundation for constructing high-quality speech understanding tasks. During the preprocessing phase, we employ audio assessment tools to score the quality of the original speech. We also apply speech enhancement models to denoise samples falling below a predefined threshold~\cite{zhao2025clearervoicestudiobridgingadvancedspeech}. Each final sample retains three modalities: video, audio, and transcript generated by the transcription models~\cite{radford2022robustspeechrecognitionlargescale,xu2025fireredasr}, ensuring readiness for subsequent multimodal input. Note that \textit{VCTK}~\cite{yamagishi2019vctk} and our internal dataset are audio-only sources, used to support accent-related tasks. Further details on data processing are provided in Appendix A~\ref{appendixA}.

\begin{table*}[!ht]
\small
\renewcommand{\arraystretch}{1.2}
\begin{tabular}{
    m{70pt} 
    m{18pt} 
    p{300pt} 
    >{\centering\arraybackslash}m{50pt} 
}
\toprule
\textbf{Task} & \textbf{Abbr.} & \textbf{Question and Option Example} & \textbf{\#(EN)} / \textbf{\#(ZH)} \\
\midrule
\multicolumn{3}{l}{\textbf{Social Attributes}} \\
\hline
   Age & Age   & Select the speaker's most likely age range. 
                  A: Adulthood  B: Adolescence... & 1057 / 507  \\
                   
 Gender & Gen   & Identify the gender of the speaker. 
                 A: Male  B: Female & 1100 / 631  \\

  Accent & Acc   & Specify the accent of the speaker. 
                  A: American B: Scottish C: Canadian... & 1000 / 1000  \\
\hline
\multicolumn{3}{l}{\textbf{Emotion Recognition}} \\
\hline
  Main Emotion& MEmo & Determine the main emotion of the speaker. A: Concerned  B: Interested... & 981 / 800  \\

  Emotions & Emos & Recognize the emotions expressed in the speaker's voice. A: Joyful B: Serious C: Bored D: Guilty E: Disgusted F: Helpless & 1000 / 799  \\
  
  Emotion Intensity & EIns & Rate the intensity of the speaker’s emotion. A: Mild B: Moderate... & 801 / 826  \\
\hline
\multicolumn{3}{l}{\textbf{Emotion Reasoning}} \\
\hline
  Emotion Shift Easy & SE & Has the speaker's emotion shifted? Yes/No & 486 / 220  \\
   Emotion Shift Hard & SH & What is the emotional change of the speaker? A: From calm to anxious B: From calm to fear C: From angry to fear D: From excited to calm & 243 / 243  \\
   Mismatch Easy & ME & Does the vocal expression contradict the speaker's actual emotion? Yes/No & 404 / 394  \\
   Mismatch Hard & MH & Identify the inconsistencies between the literal meaning of the speaker's lines and their actual emotion. A: Literal: Amused, actual: Surprised. B: Literal: Sad, actual: Proud C: Literal: Happy, actual: Disappointed ...
 & 347 / 197  \\
\hline
\multicolumn{3}{l}{\textbf{Nonverbal Behavior}} \\
\hline
   Speech Style & Style & Specify the speech style of the primary speaker. A: Questioning B: Shouting C: Neutral D: Speaking with a quiet smile & 767 / 386 \\
   Visual & Vis  & Guess the speaker's visible behavior. A: Shouts B: Cries C: Laughs ... & 742 / 665  \\
   Description & Desc & Is the following description of the speaker accurate? A lady whispers, with wind blowing her hair, raised eyebrows and lip corners.  Yes/No & 1250 / 631\\
\hline
\multicolumn{3}{l}{\textbf{Discourse Context}} \\
\hline
   Intent & Intent & Infer the speaker's intention. A: Command B: Suggest C: Inform ... & 549 / 724  \\
   Scene  & Scn  & Determine the conversation scene being depicted. A: Cooperative B: Directive C: Inquisitive D: Narrative E: Confrontational F: Avoidant & 690 / 698 \\
   Subtext & Subt  & Does the following description accurately reflect the speaker's underlying intention? The speaker is being sarcastic, implying the situation is not actually good. Transcript: Well, the good news is, so many.... Yes/No & 163 / 65\\
\midrule
\textbf{Total} & & & \textbf{11580} / \textbf{8786} \\
\bottomrule
\end{tabular}
\caption{Benchmark overview—task abbreviations, examples, and item counts.}
\label{tab:task_overview}
\end{table*}

\textbf{Step 2: Information extraction and cross-validation.} Inspiring by human multimodal perception and reasoning in daily communication, we designed a multi-level process for information extraction and validation. First, we input the transcript into an \textit{LLM} to infer several possible speaker states, such as different emotional states or communicative intents. They are summarized and used as prior knowledge. We then combine these inferred attributes with corresponding audio and visual modalities and input them into audio and vision models, respectively, in order to extract each modality's representation of the speaker's state. To enhance the consistency and reliability of multimodal information, we employ a single-modality cross-validation strategy. That is, the visual description is fed into the audio model, and vice versa. That can assess the compatibility and logical coherence of content generated by each modality. In this way, we obtain complementary visual and auditory semantic descriptions, providing multi-perspective support for subsequent information fusion.

\textbf{Step 3: Information fusion and expert verification.}
We employ a hierarchical fusion strategy managed by a \textit{QWQ} inference model \cite{qwen2025qwen25technicalreport}. This process synthesizes the validated information into a multi-dimensional open-ended description for each audio sample, structured across three semantic layers: (1) a core layer used to capture dominant attributes like emotion or intent; (2) a detail layer for fine-grained supporting cues (e.g., ``smile''); and (3) a background layer for contextual elements.

Afterward, we divide these labeled data as follows: 80\% of this data constitutes the \textbf{H}uman-level \textbf{P}erception \textbf{S}poken Speech \textbf{C}aption (\textit{HPSC}) dataset, comprising approximately 50,000 speech-caption pairs. The \textit{HPSU} benchmark is derived from the remaining, distinct 20\% subset. For this portion, we first utilize \textit{Gemini 2.5 Pro} \cite{comanici2025gemini25pushingfrontier} to structure the multi-layer descriptions into discrete evaluation triplets. These triplets then undergo a stringent human verification protocol: three trained annotators independently review each instance. Only those achieving unanimous agreement are retained. This rigorous process yields a final acceptance rate of 81.26\%, ensuring the high fidelity and reliability of the \textit{HPSU} benchmark.

We release the 50,000-instance \textit{HPSC} dataset to catalyze further research. For empirical validation, we conduct supervised fine-tuning (\textit{SFT}) on open-source \textit{Speech LLMs}. The results, detailed in Appendix B~\ref{sec:sft}, showed that tuning on \textit{HPSC} data can enhance a model's perceptual and comprehension capabilities. Besides, \textit{HPSC} holds considerable potential for future applications in areas such as controllable speech generation.

\subsection{HPSU Benchmark}

\subsubsection{Task Design}

The \textit{HPSU} benchmark is architected to evaluate the nuanced comprehension abilities of \textit{Speech LLMs}. Our framework is organized hierarchically across two levels of cognitive complexity: basic perception and complex inference, encompassing 5 distinct domains and 16 unique tasks, as detailed in Table~\ref{tab:task_overview}. This structure can effectively evaluate the model, from recognizing speaker attributes to inferring deep, context-dependent meaning.

We operationalize our tasks using varied question formats to enhance comparability and constrain the open-ended nature of \textit{LLM} outputs. Objective tasks are formulated as single-choice questions for precise judgment. More complex phenomena, such as identifying concurrent emotions, are tested using multiple-choice formats. Deep semantic deviation, like subtext interpretation, is assessed via Yes/No questions. For subjective tasks, we employ a sophisticated distractor-generation mechanism based on \textit{Gemini 2.5 Pro}. It can create semantically calibrated distractors (``similar'', ``middle'', ``opposite''), so as to increase the challenge and probe the model's fine-grained discriminative power.

To explicitly measure robustness, we integrated an adversarial induction protocol directly into the benchmark's design. Each relevant task instance includes three prompt variations: a standard unbiased query, a positive prompt with answer-leaking hints, and a negative prompt with misleading cues. As shown in Table~\ref{tab:prompt types}, multiple templates are used for each type to mitigate prompt sensitivity. This multi-faceted design ensures \textit{HPSU} provides a robust, in-depth, and fair evaluation of human-level speech understanding.

\begin{table}[ht]
\centering
\setlength{\tabcolsep}{4pt}
\begin{tabular} {m{40pt} m{180pt}}
\hline
\textbf{Prompt Type} & \textbf{Example (Main Emotion task)} \\
\hline
Standard & Please identify the speaker's main emotion.\\
\hline
Positive & The main emotion of the speaker seems to be happy. Is this judgment correct? Please identify the speaker's main emotion. \\
\hline
Negative & The main emotion of the speaker seems to be upset. Is this judgment correct? Please identify the speaker's main emotion. \\
\hline
\end{tabular}
\caption{Example of prompt induction in the Main Emotion task (Note: In this example, "happy" is the correct answer).}
\label{tab:prompt types}
\end{table}

\subsubsection{Comparison with Other Benchmarks}

\begin{table}[ht]
\centering
\setlength{\tabcolsep}{4pt}
\begin{tabular}{lccccc}
\hline
\textbf{Abbr.} & \textbf{MMAU} & \textbf{AIR} & \textbf{MMAR} & \textbf{MMSU} & \textbf{HPSU} \\
\hline
num        & 3k+   & 9k+   & 0.6k+   & 5k   & 20k+ \\
Age                        & \cmark & \cmark & \xmark & \cmark & \cmark \\
Gen                     & \cmark & \cmark & \xmark & \cmark & \cmark \\
Acc                     & \cmark & \ocircle & \cmark & \cmark & \cmark \\
MEmo                    & \ocircle & \ocircle & \ocircle & \ocircle & \cmark \\
Emos                   & \ocircle & \xmark & \xmark & \xmark & \cmark \\
EIns                  & \xmark & \xmark & \xmark & \xmark & \cmark \\
SE       & \cmark & \xmark & \xmark & \xmark & \cmark \\
SH             & \ocircle& \xmark & \xmark & \xmark & \cmark \\
ME             & \xmark & \xmark & \xmark & \xmark & \cmark \\
MH             & \xmark & \xmark & \xmark & \xmark & \cmark \\
Style              & \xmark & \xmark & \ocircle & \xmark & \cmark \\
Vis                   & \xmark & \xmark & \ocircle & \xmark & \cmark \\
Desc                & \xmark & \xmark & \xmark & \xmark & \cmark \\
Intent                & \cmark & \cmark & \ocircle & \cmark & \cmark \\
Scn                      & \xmark & \xmark & \xmark & \xmark & \cmark \\
Subt                    & \xmark & \xmark & \xmark & \xmark & \cmark \\
\hline
\end{tabular}
\caption{Transposed comparison of speech-related benchmarks. \cmark: covered; 
\ocircle: partially related but with a different focus or at a different granularity; 
\xmark: not covered.}
\label{tab:comparison}
\end{table}

As summarized in Table~\ref{tab:comparison}, \textit{HPSU} has three contributions compared to previous works. Firstly, its scale of over 20,000 expert-validated instances provides a foundation for statistically robust evaluation, surpassing other benchmarks. Secondly, it introduces a wide range of in-depth tasks. These tasks have challenges such as multi-emotion tracking and implicit emotion inference that are designed to probe a model's ability to handle the complex and dynamic information in real-world speech. Finally, by incorporating both English and Chinese, \textit{HPSU} facilitates crucial multilingual analysis of model performance across different linguistic and cultural contexts.

\subsubsection{Evaluation Strategy}

We prioritize semantic accuracy over superficial lexical matching, using \textit{Gemini 2.5 Pro} as an automated adjudicator to ensure fair and consistent scoring across diverse model outputs. The primary scoring mechanism is strict: for Single-Choice and Judgment/Binary tasks, only exact semantic equivalence with the reference answer receives credit. For Multiple-Choice tasks, any false positive results in a score of zero, while partial credit is awarded only for omissions in the absence of incorrect selections.

Beyond this primary scoring, our strategy leverages the benchmark's unique design for deeper analysis. First, we conduct a quantitative analysis using the graded answer options. Specifically, we analyze the distribution of model responses across the ``true'', ``similar'', ``middle'', and ``opposite'' tiers to understand its decision-making process under uncertainty. Second, we assess model robustness by comparing performance across the three induction prompt types. This allows us to quantify a model's stability and its resistance to conversational biases and misleading information, addressing a critical gap in prior evaluation methodologies.

\section{Results and Discussion}

\begin{table*}[!ht]
\small
\centering
\setlength{\tabcolsep}{3pt}
\renewcommand{\arraystretch}{1.2}
\begin{tabular}{@{} l c *{5}{c} @{}}
\toprule
\textbf{Lang} & \textbf{Models} 
& \textbf{Social Attributes} 
& \textbf{Emotion Recognition} 
& \textbf{Emotion Reasoning} 
& \textbf{Nonverbal Behavior} 
& \textbf{Discourse Context} \\
\cmidrule(lr){3-7}
& & Age \textbar{} Gen \textbar{} Acc
& MEmo \textbar{} Emos \textbar{} EIns
& SE \textbar{} SH \textbar{} ME \textbar{} MH
& Style \textbar{} Vis \textbar{} Desc
& Intent \textbar{} Scn \textbar{} Subt \\
\midrule

\multirow{20}{*}[3em]{EN} 
& Human
& 73.3 \textbar{} 99.8 \textbar{} 67.9
& 85.4 \textbar{} 81.9 \textbar{} 89.1
& 81.9 \textbar{} 78.6 \textbar{} 79.0 \textbar{} 85.0
& 84.2 \textbar{} 90.3 \textbar{} 94.9
& 86.1 \textbar{} 85.3 \textbar{} 89.4 \\

& Random Guess
& 25.9 \textbar{} 50.6 \textbar{} 10.3
& 25.2 \textbar{} 5.0 \textbar{} 31.2
& 50.1 \textbar{} 26.8 \textbar{} 45.8 \textbar{} 25.4
& 24.8 \textbar{} 25.9 \textbar{} 50.5
& 24.6 \textbar{} 16.7 \textbar{} 49.9 \\

& Whisper+GPT 
& 43.3 \textbar{} 34.6 \textbar{} 23.8
& 58.1 \textbar{} 70.1 \textbar{} 51.0
& 51.9 \textbar{} 35.1 \textbar{} 51.2 \textbar{} 57.4
& 49.4 \textbar{} 46.3 \textbar{} 49.0
& 78.5 \textbar{} 44.6 \textbar{} 66.4 \\

\cline{2-7}

& AF2 
& 22.8 \textbar{} 90.1 \textbar{} 16.8
& 10.9 \textbar{} 25.1 \textbar{} 25.8
& 45.7 \textbar{} 28.2 \textbar{} 30.1 \textbar{} 30.8
& 10.9 \textbar{} 20.8 \textbar{} 60.8
& 13.8 \textbar{} 20.7 \textbar{} 62.8 \\

& AF3 
& 46.6 \textbar{} 86.8 \textbar{} 21.7
& 54.9 \textbar{} 53.0 \textbar{} 56.5
& 58.2 \textbar{} \textbf{57.0} \textbar{} 50.2 \textbar{} 48.3
& 48.1 \textbar{} 55.4 \textbar{} 61.6
& 67.4 \textbar{} 39.9 \textbar{} 65.9 \\

& Kimi-Audio 
& 55.0 \textbar{} 92.0 \textbar{} 47.4
& \textbf{62.7} \textbar{} 46.9 \textbar{} 49.1
& 55.2 \textbar{} 41.4 \textbar{} 53.3 \textbar{} 41.8
& \textbf{64.8} \textbar{} \textbf{63.0} \textbar{} 62.6
& 61.7 \textbar{} 36.2 \textbar{} 69.5 \\

& Qwen-Audio 
& 17.9 \textbar{} 55.6 \textbar{} 30.2
& 33.7 \textbar{} 27.1 \textbar{} 18.9
& 13.5 \textbar{} 32.6 \textbar{} 19.3 \textbar{} 32.7
& 40.7 \textbar{} 21.4 \textbar{} 51.8
& 44.0 \textbar{} 19.9 \textbar{} 66.5 \\

& Qwen2-Audio
& 37.6 \textbar{} 89.4 \textbar{} 28.4
& 53.4 \textbar{} 39.4 \textbar{} 47.0
& 51.4 \textbar{} 44.1 \textbar{} 49.6 \textbar{} 57.1
& 34.6 \textbar{} 50.7 \textbar{} 59.0
& 60.6 \textbar{} 30.8 \textbar{} 59.8 \\

& SALMONN
& \textbf{61.0} \textbar{} 79.0 \textbar{} 0.0
& 38.2 \textbar{} 26.9 \textbar{} 29.2
& 51.1 \textbar{} 32.9 \textbar{} 49.8 \textbar{} \textbf{64.9}
& 22.1 \textbar{} 23.9 \textbar{} 58.1
& 23.8 \textbar{} 27.9 \textbar{} 42.7 \\

& Soundwave
& 40.3 \textbar{} 75.5 \textbar{} 19.2
& 51.9 \textbar{} 23.2 \textbar{} 9.5
& 52.4 \textbar{} 39.2 \textbar{} 52.5 \textbar{} 45.1
& 59.5 \textbar{} 50.8 \textbar{} 50.9
& 55.1 \textbar{} 34.6 \textbar{} 50.1 \\

\cline{2-7}

& BC-Omni-1.5
& 16.9 \textbar{} 53.7 \textbar{} 28.0
& 48.2 \textbar{} 45.5 \textbar{} 45.7
& 51.8 \textbar{} 33.3 \textbar{} 25.9 \textbar{} 41.0
& 53.5 \textbar{} 44.8 \textbar{} 54.1
& 57.1 \textbar{} 42.3 \textbar{} 68.9 \\

& Qwen-2.5-Omni
& 54.7 \textbar{} \textbf{94.6} \textbar{} 23.6
& 49.4 \textbar{} 56.8 \textbar{} 52.9
& \textbf{60.9} \textbar{} 47.9 \textbar{} 55.0 \textbar{} 51.0
& 54.5 \textbar{} 57.4 \textbar{} 57.5
& 69.3 \textbar{} 49.4 \textbar{} 72.5 \\

& Gemini 2.5 Flash
& 44.9 \textbar{} 88.6 \textbar{} 37.0
& 53.6 \textbar{} 69.8 \textbar{} 58.9
& 49.3 \textbar{} 39.3 \textbar{} 53.6 \textbar{} 38.1
& 58.9 \textbar{} 44.7 \textbar{} 53.1
& 77.3 \textbar{} 45.0 \textbar{} 66.8 \\

\cline{2-7}

& AF3.5
& 50.1 \textbar{} 81.5 \textbar{} 20.0
& 54.7 \textbar{} 45.6 \textbar{} 47.0
& 52.1 \textbar{} 43.2 \textbar{} 52.6 \textbar{} 31.1
& 46.2 \textbar{} 51.0 \textbar{} \textbf{78.1}
& 61.5 \textbar{} 43.1 \textbar{} \textbf{75.6} \\

& Audio-Reasoner
& 30.4 \textbar{} 67.8 \textbar{} 30.5
& 59.1 \textbar{} 49.2 \textbar{} 43.0
& 59.7 \textbar{} 49.3 \textbar{} 56.5 \textbar{} 50.0
& 61.3 \textbar{} 56.9 \textbar{} 59.9
& 69.5 \textbar{} 46.1 \textbar{} 68.1 \\

& Gemini 2.5 Pro 
& 60.4 \textbar{} 93.1 \textbar{} \textbf{49.9}
& 51.4 \textbar{} \textbf{72.0} \textbar{} \textbf{59.6}
& 59.0 \textbar{} 49.7 \textbar{} \textbf{56.8} \textbar{} 54.3
& 53.5 \textbar{} 52.2 \textbar{} 51.6
& \textbf{79.4} \textbar{} \textbf{48.0} \textbar{} 59.1 \\

\midrule

\midrule
\multirow{20}{*}[5em]{ZH} 
& Human
& 86.9 \textbar{} 98.2 \textbar{} 70.2
& 95.5 \textbar{} 84.5 \textbar{} 95.4 
& 82.1 \textbar{} 85.6 \textbar{} 89.6 \textbar{} 88.4
& 91.9 \textbar{} 95.9 \textbar{} 97.4
& 91.7 \textbar{} 93.2 \textbar{} 95.9 \\

& Random Guess
& 26.0 \textbar{} 45.3 \textbar{} 10.5
& 25.1 \textbar{} 4.5 \textbar{} 33.1 
& 53.1 \textbar{} 26.3 \textbar{} 50.6 \textbar{} 26.3
& 26.9 \textbar{} 25.6 \textbar{} 51.1
& 26.0 \textbar{} 16.3 \textbar{} 50.8 \\

& Whisper+GPT
& 39.7 \textbar{} 54.8 \textbar{} 12.7
& 46.4 \textbar{} 47.2 \textbar{} 54.6
& 67.5 \textbar{} 55.9 \textbar{} 60.8 \textbar{} 53.5
& 50.7 \textbar{} 74.0 \textbar{} 51.9
& 79.3 \textbar{} 62.5 \textbar{} 53.8 \\

\cline{2-7}

& AF3
& 41.4 \textbar{} 93.4 \textbar{} 14.6
& 49.8 \textbar{} 12.5 \textbar{} 36.9
& 46.8 \textbar{} 55.0 \textbar{} 48.5 \textbar{} 54.2
& 48.3 \textbar{} 60.1 \textbar{} 50.1
& 58.6 \textbar{} 42.7 \textbar{} 56.9 \\

& Kimi-Audio
& \textbf{56.8} \textbar{} 78.6 \textbar{} 19.0
& 50.8 \textbar{} 30.5 \textbar{} 47.5
& \textbf{67.8} \textbar{} 46.9 \textbar{} 39.4 \textbar{} 45.1
& 61.1 \textbar{} 70.0 \textbar{} 65.8
& 59.0 \textbar{} 45.2 \textbar{} 56.9 \\

& Qwen-Audio
& 38.9 \textbar{} 79.3 \textbar{} \textbf{53.2}
& 31.0 \textbar{} 6.0 \textbar{} 26.7
& 32.1 \textbar{} 30.7 \textbar{} 27.9 \textbar{} 38.0
& 38.1 \textbar{} 27.4 \textbar{} 52.0
& 40.3 \textbar{} 18.0 \textbar{} 49.2 \\

& Qwen2-Audio
& 35.0 \textbar{} 93.2 \textbar{} 24.5
& 59.4 \textbar{} 32.8 \textbar{} 47.4
& 51.5 \textbar{} 52.4 \textbar{} 53.2 \textbar{} 54.8
& 60.7 \textbar{} 52.6 \textbar{} 62.6
& 74.4 \textbar{} 40.9 \textbar{} 56.9 \\

& SALMONN
& 23.8 \textbar{} 61.7 \textbar{} 0.1
& 27.1 \textbar{} 8.8 \textbar{} 38.3
& 57.0 \textbar{} 45.6 \textbar{} \textbf{67.2} \textbar{} 64.0
& 51.2 \textbar{} 36.0 \textbar{} 47.1
& 34.1 \textbar{} 26.9 \textbar{} 50.8 \\

\cline{2-7}

& BC-Omni-1.5
& 27.1 \textbar{} 45.5 \textbar{} 8.0
& 32.9 \textbar{} 24.5 \textbar{} 35.7
& 34.4 \textbar{} 44.1 \textbar{} 49.2 \textbar{} 42.6
& 37.0 \textbar{} 36.4 \textbar{} 43.0
& 60.6 \textbar{} 41.2 \textbar{} 61.5 \\

& Qwen-2.5-Omni
& 53.3 \textbar{} \textbf{97.9} \textbar{} 38.0
& \textbf{60.1} \textbar{} 40.0 \textbar{} 57.6
& 60.1 \textbar{} 76.3 \textbar{} 45.7 \textbar{} 58.2
& \textbf{72.7} \textbar{} \textbf{85.0} \textbar{} \textbf{79.2}
& 78.1 \textbar{} 59.9 \textbar{} 49.2 \\

& Gemini 2.5 Flash
& 49.2 \textbar{} 92.4 \textbar{} 15.3
& 57.8 \textbar{} 45.9 \textbar{} 62.4
& 40.5 \textbar{} 53.7 \textbar{} 58.2 \textbar{} 55.8
& 60.7 \textbar{} 54.2 \textbar{} 67.2
& 74.4 \textbar{} 62.3 \textbar{} 62.9 \\

\cline{2-7}

& Audio-Reasoner
& 32.9 \textbar{} 76.5 \textbar{} 14.6
& 54.5 \textbar{} 23.9 \textbar{} 50.0
& 57.4 \textbar{} 65.7 \textbar{} 49.8 \textbar{} 59.9
& 56.6 \textbar{} 60.4 \textbar{} 62.8
& 73.5 \textbar{} 56.1 \textbar{} \textbf{63.1} \\

& Gemini 2.5 Pro
& 55.4 \textbar{} 97.0 \textbar{} 25.3
& 58.6 \textbar{} \textbf{69.0} \textbar{} \textbf{66.1}
& 59.7 \textbar{} \textbf{80.5} \textbar{} 61.2 \textbar{} \textbf{76.0}
& 71.4 \textbar{} 74.2 \textbar{} 64.0
& \textbf{80.2} \textbar{} \textbf{62.6} \textbar{} 52.3 \\

\bottomrule
\end{tabular}
\caption{Performance comparison of different models on \textit{HPSU}. All values represent accuracy($\uparrow$). For each single-language task, the best-performing model’s accuracy is highlighted in bold. \textit{AF2}, \textit{AF3.5}, and \textit{Soundwave} are excluded from the ZH section as they do not support our tasks in Chinese.}
\label{tab:main_results}
\end{table*}

We conducted our evaluation on 13 models, comprising 11 leading open-source systems (\textit{Audio Flamingo 2}~\cite{ghosh2025audio}, \textit{Audio Flamingo 3}~\cite{goel2025audio}, \textit{Kimi-Audio-Instruct}~\cite{ding2025kimi}, \textit{Qwen-Audio-Chat}~\cite{chu2023qwen}, \textit{Qwen2-Audio-Instruct}~\cite{chu2024qwen2}, \textit{SALMONN}~\cite{tang2024salmonn}, \textit{Soundwave}~\cite{zhang2025soundwave}, \textit{Baichuan-Omni-1.5}~\cite{li2025baichuan}, \textit{Qwen2.5-Omni}~\cite{xu2025qwen2.5omni}, \textit{Audio Flamingo 3.5}~\cite{goel2025audio}, \textit{Audio-Reasoner}~\cite{zhifei2025audioreasoner}) and 2 proprietary models (\textit{Gemini 2.5 Flash and Pro}). Notably, the native \textit{GPT-4o} audio model \cite{openai2024gpt4technicalreport} was excluded from our comparison as it refused to answer a majority of our queries, citing safety policies.

We established three baselines to compare the model's performance. The \textit{Human} baseline, an empirical performance ceiling, was derived from 10 native speakers per language answering 300 stratified-sampled items. A \textit{Random Guess} baseline provided a chance-level reference. Finally, a \textit{Whisper+GPT} cascade, which fed \textit{Whisper} transcriptions into \textit{GPT-4o} was included to isolate the performance gains attributable to the acoustic modality beyond mere text. Details for all evaluated models and baselines are provided in Appendix C.

\subsection{Main Results}

\begin{figure}[!h]
\centering
\includegraphics[width=0.48\textwidth,keepaspectratio]{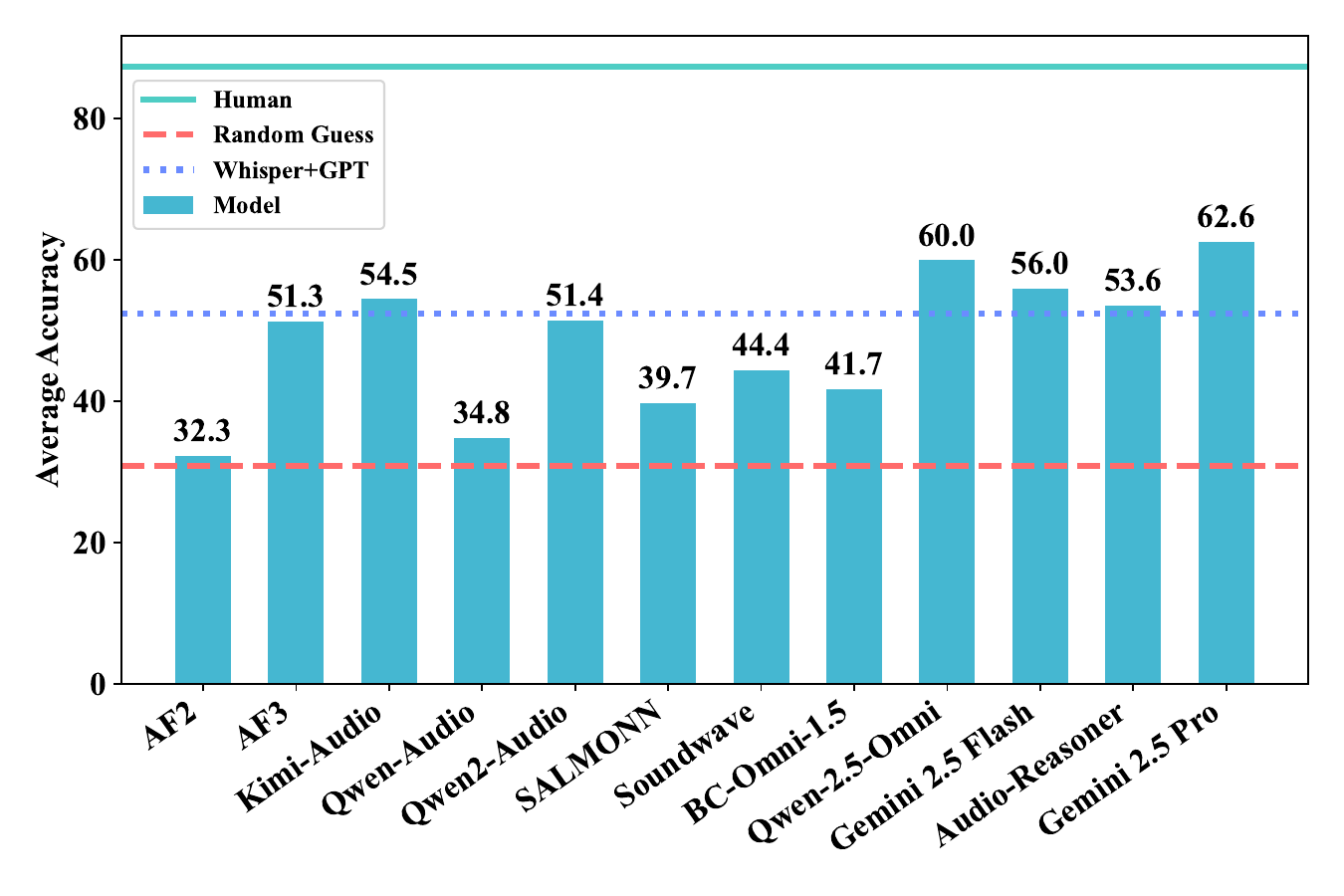}
\caption{Overall accuracy of different models.}
\label{fig:average accuracy}
\end{figure}

As presented in Table~\ref{tab:main_results} and Figure~\ref{fig:average accuracy}, the evaluation results validated the difficulty of \textit{HPSU}. To a certain extent, they reflected the capabilities and limitations of current \textit{Speech LLMs}.

\textbf{Challenging Nature of the HPSU Benchmark.} The \textit{HPSU} benchmark posed considerable challenges to current speech understanding models. Human evaluators achieved an average accuracy of 87.3\%, whereas the best-performing model, \textit{Gemini 2.5 Pro}, only obtained an accuracy of 62.6\%. This substantial performance gap underscored the disparity between human speech-understanding capabilities and those of current models. That reflected the rigor of the \textit{HPSU} benchmark, and simultaneously highlighted potential opportunities for future research.

\textbf{Differential Performance Across Task Types.} Detailed analysis revealed distinct performance discrepancies across task categories. Specifically, models performed nearly at human levels in basic recognition tasks, such as gender identification, but underperformed in high-level semantic reasoning tasks like \textit{Scn} and \textit{MH}. These results indicated the considerable challenges faced by current models in deeply comprehending complex semantic nuances in speech.

\textbf{Competitive Performance Between Open-source and Proprietary Models.} Notably, there is a minimal performance gap between leading open-source models and proprietary models. For example, the open-source model \textit{Qwen2.5-Omni} achieved an average accuracy of 60.0\%, only 2.6\% behind the proprietary \textit{Gemini 2.5 Pro} model, and even surpassed \textit{Gemini 2.5 Pro} on tasks of \textit{SE}(EN) and \textit{Vis}(ZH). That indicated in complex speech understanding tasks, if the training data is limited and diverse, the performance gap will narrow, enabling open-source models and proprietary models to achieve good performance.

\textbf{Performance Analysis of Cascade Approaches.} Additionally, cascade approaches often used ``speech transcription followed by large language models'' (\textit{Whisper + GPT-4o}). They obtained competitive performance with \textit{Gemini 2.5 Pro} on \textit{Emos} and \textit{Intent} tasks. However, they performed worse in those tasks that heavily relied on acoustic details, such as \textit{Desc} and \textit{SH}. This performance disparity indicated the robust semantic comprehension and generalization capability of the textual modality. It accentuated the indispensability of acoustic information. Also, it reflected the need for advanced multimodal fusion strategies and richer training datasets.

\textbf{Impact of Audio Quality on Human Performance.} Data analysis indicated that human evaluators consistently achieved higher scores on Chinese tasks compared to English tasks. That may be due to variations in audio quality from different data sources. Nevertheless, performance trends remained consistent across languages, reflecting inherent task difficulties rather than linguistic factors. Thus, human performance metrics served as important benchmarks for evaluating models' performance across tasks.

\subsection{Weaknesses in Speech LLMs}
\label{subsec:weakness_analysis}

The performance gap between basic and deep speech understanding models revealed a systematic bias in their training data. While \textit{Speech LLMs} demonstrated strong capabilities on basic perceptual tasks, they failed to emulate human ability to understand complex, real-world speech. We believed that this data bias was the cause. This bias originated from the corpora used for pre-training. Models were overwhelmingly trained on large-scale datasets that heavily skewed towards ``low-level'' tasks like \textit{ASR} or emotion classification. That reflected they performed well in these areas, but were unable to handle the \textit{higher-order} cognitive tasks that involved potential intentions or subtle emotional changes.

The reason for this data imbalance lied in the fact that it was difficult to label these higher-order tasks. This not only involved high costs and a long time consumption, but also required high cognitive abilities. Thus, the scarcity of appropriate training data would lead to the observed deficiencies in the advanced reasoning abilities of current models. This reflected the guiding value of \textit{HPSU}. Our benchmark can quantify this \textit{higher-order capability gap}, establishing a clear target to steer the community from the over-optimization of basic tasks toward the strategic exploration of human-level understanding. 

\begin{figure}
    \centering
    \includegraphics[width=0.48\textwidth,keepaspectratio]{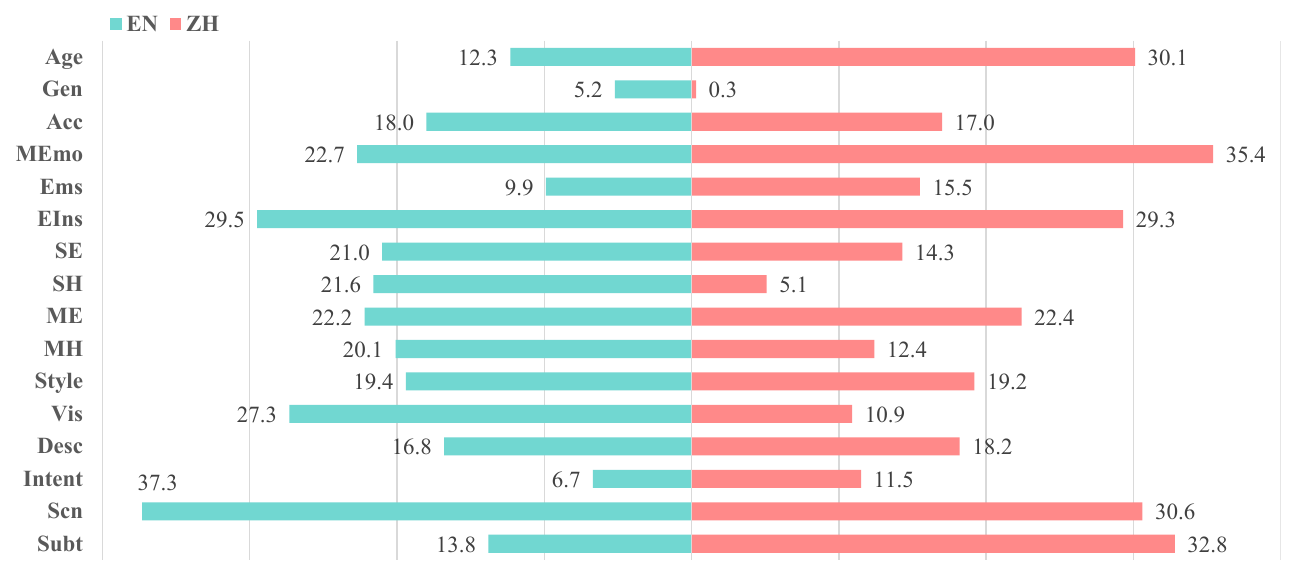}
    \caption{Difference between the best results of the models and those of human.}
    \label{fig:model_human_gap}
\end{figure}

Figure \ref{fig:model_human_gap} quantified the gap between the best results of the models and those of humans. The models not only underperformed humans on most tasks but also exhibited large cross-lingual performance gaps within the same task. This situation was not observed in humans. For example, on the task of subtext interpretation, some models performed worse than humans in English, but their performance in Chinese was inferior to that of humans. That implied grasping Chinese subtext was challenging for current models. The reason may be that this expressive form was more elaborate and subtle. Thus, a multilingual benchmark is important to fully expose the capabilities and limitations of \textit{Speech LLMs}.

\subsection{Model Susceptibility to Induction}
\label{subsec:induction_analysis}

\begin{figure}[!ht]
\centering
\includegraphics[width=0.45\textwidth,keepaspectratio]{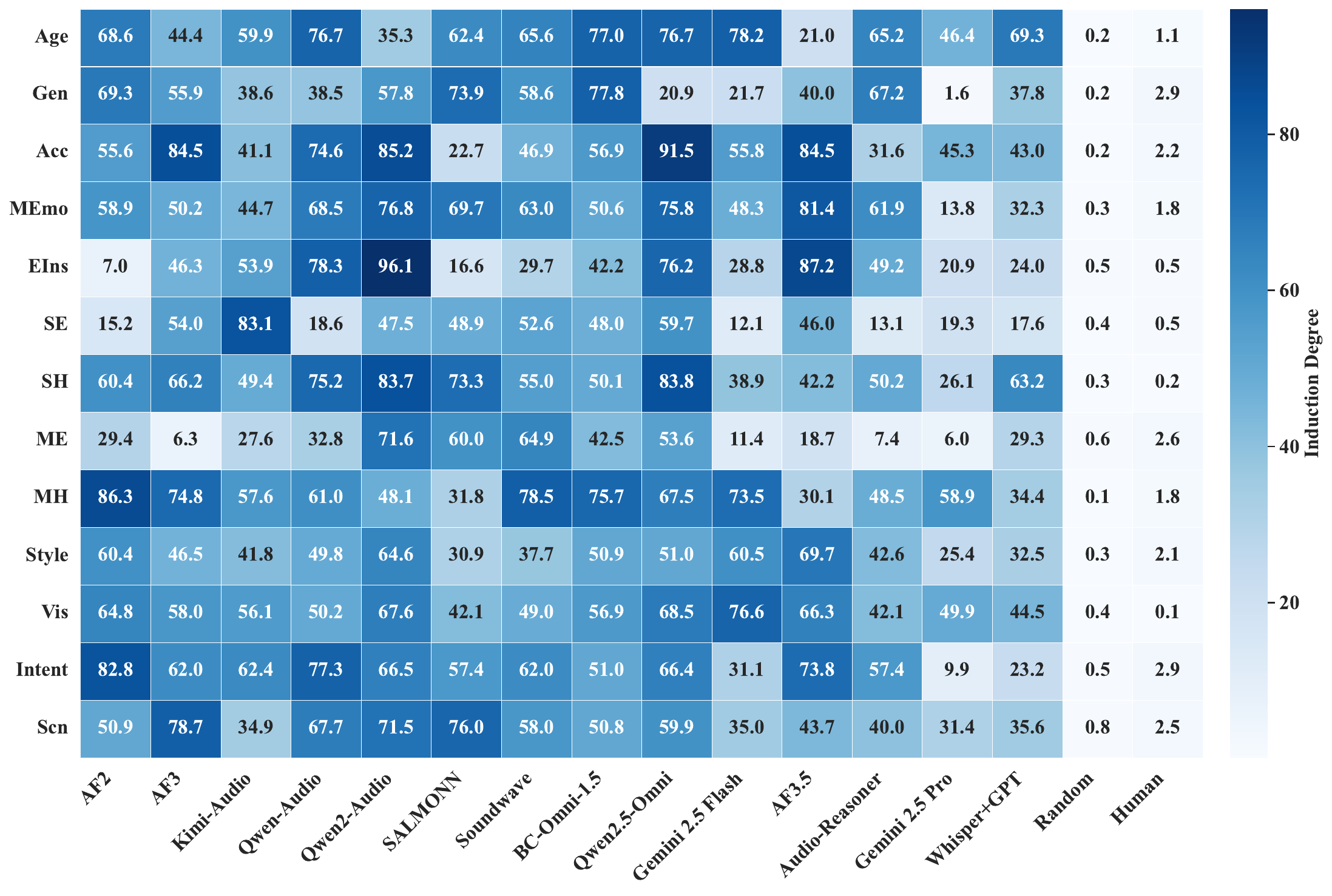}
\caption{Susceptibility of models to the biased prompts. Darker shades indicate a higher degree of susceptibility.}
\label{fig:induction_heatmap}
\end{figure}

Our adversarial evaluation revealed the flaws in current \textit{Speech LLMs} when dealing with misleading information.  But these flaws do not occur in humans. As shown in Figure \ref{fig:induction_heatmap}, we calculated the performance spread across Standard, Positive, and Negative prompts for each model and each task, where the prompts were displayed in Table~\ref{tab:prompt types}. Each cell in the heatmap indicated the difference between the best and worst results across prompt variants. As presented in Figure \ref{fig:induction_heatmap}, human evaluators were almost immune to induced prompts. In contrast, all evaluated models exhibit a high degree of susceptibility, as their judgments were easily swayed by incorrect cues embedded in the prompt. This indicated there was a gap between the humans' cognitive filtering and the vulnerability of model reasoning.

This vulnerability was not uniform, concentrating in two task categories. Models were particularly fragile on tasks with high intrinsic ambiguity (e.g., \textit{Age}, \textit{Acc}, \textit{MH}), where high uncertainty appeared to make them default to prompt cues over audio evidence. They were more susceptible when dealing with open-ended tasks, i.e., tasks with a wide range of possible answers like \textit{Vis}, \textit{Intent}, \textit{Scn}. A single low-frequency cue in the prompt may draw the model's attention, thereby hindering its analysis of the speech content.

The architectural trend was correlated with robustness. Models with stronger reasoning ability and those with native omni-modal designs (such as \textit{Gemini}) exhibited greater resilience. That revealed fine-tuning of a text-centric \textit{LLM} would introduce \textit{textual bias}, making it susceptible to being manipulated by the prompts' content. Conversely, architectures designed for equitable multimodal integration appear less dependent on textual cues. Thus, it had better robustness against adversarial induction.

\begin{figure}[!h]
\centering
\includegraphics[width=0.45\textwidth,keepaspectratio]{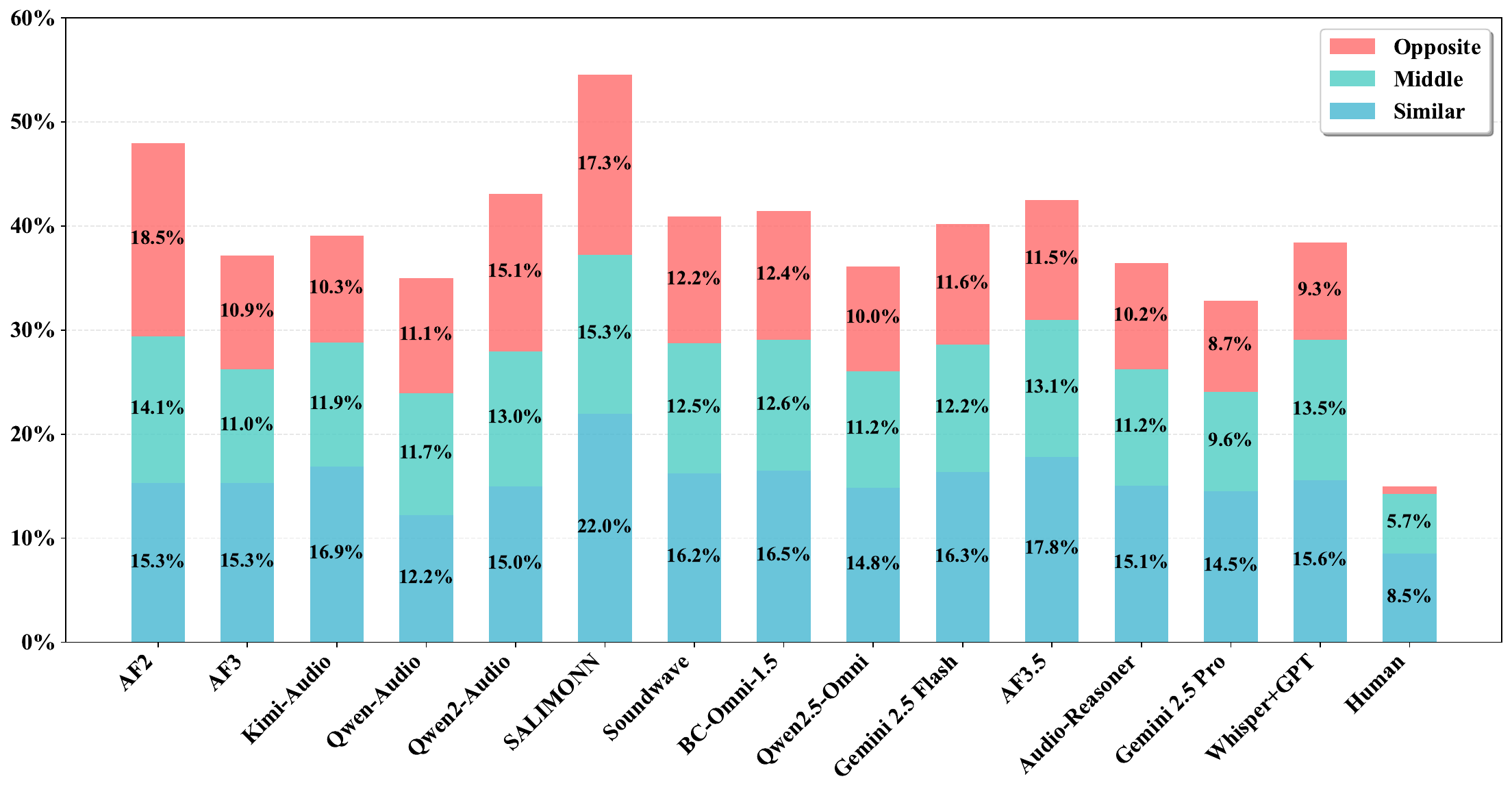}
\caption{Error type distributions across models.}
\label{fig:error type}
\end{figure}

\subsection{Analysis of Graded Evaluation}
\label{subsec:graded_eval_analysis}

The graded evaluation strategy provided a high-fidelity diagnostic tool. It was no longer confined to simple binary scoring but capable of capturing the subtle differences in the model's reasoning process during ambiguous tasks. As shown in Figure \ref{fig:error type}, we observed the cause of the model failure lied in the lack of fine-grained discrimination, rather than the loss of semantic coherence. Models frequently select ``Similar'' options, indicating partial understanding, but they always avoid choosing the ``Opposite'' ones. This suggested the challenge is not a lack of comprehension ability, but rather the inability to solve those subtle and ambiguous problems that humans can easily handle. Besides, this framework enables a more nuanced comparison between models, revealing subtle differences in their ability to resolve ambiguity. For example, while both \textit{Gemini 2.5 Pro} and \textit{Qwen-2.5-Omni} perform strongly, \textit{Gemini} exhibited a slightly higher ratio of ``True'' to ``Similar'' responses. That revealed a better capability in resolving ambiguity. Furthermore, the non-zero ``Similar'' score from human evaluators validated the inherent subjectivity of these tasks. That revealed the task's true goal was not objective accuracy but human-like precision in navigating ambiguity.

\section{Related Work}


The advent of \textit{Speech LLMs} has enhanced audio understanding, moving from cascaded systems that lose crucial acoustic information to end-to-end architectures. This new generation includes \textit{Qwen-Audio} series, \textit{Kimi-Audio-Instruct}, \textit{SALMONN}, \textit{Soundwave}, and \textit{Audio Flamingo}, which directly integrate audio with \textit{LLMs}. The scope has further expanded with \textit{Omni-Language Models} such as \textit{Baichuan-Omni} and \textit{Qwen-2.5-Omni} that process multiple modalities, and \textit{Large Audio Reasoning Models} like \textit{Audio-reasoner}, \textit{Audio Flamingo 3.5}, and \textit{Gemini 2.5} that are specialized for complex inference. However, the evaluation of these powerful models has not kept pace, remaining fragmented and focused on surface-level tasks. There is still a gap in the assessment of the deep perception and inference required to understand implicit intent, emotional dynamics, and real-world pragmatic context. \textit{HPSU} is designed to fill this gap.


The evaluation of audio understanding has evolved from basic benchmarks for discrete acoustic events (\textit{TUT Sound Events} \cite{mesaros2016tut}) and scene classification (\textit{CochlScene} \cite{jeong2022cochlscene}), to frameworks assessing the generalizability of pretrained representations like \textit{SUPERB} \cite{yang2021superb} and \textit{HEAR} \cite{turian2022hear}. The exploration of deeper comprehension continues. The benchmarks for audio question-answering (\textit{ClothoAQA} \cite{lipping2022clotho}) and complex reasoning (\textit{MMAU} \cite{sakshi2025mmau}) have been established. Recently, \textit{Large Audio Models} spurred benchmarks with interactive capabilities, such as instruction following (\textit{AIR-Bench} \cite{yang2024air}), open-ended evaluation (\textit{AudioBench} \cite{wang2025audiobench}), and paralinguistic understanding (\textit{SD-eval} \cite{ao2024sdeval}). Notably, \textit{MMSU} \cite{wang2025mmsu} integrates linguistic theory to evaluate fine-grained perception in spoken language, while \textit{MMAR} \cite{ma2025mmar} assesses deep reasoning across a broad spectrum of real-world audio scenarios. However, there is still a gap. These benchmarks overlook the nuanced, implicit, and dynamic comprehension required for human interaction. This human ability often relies on non-conversational or synthetic data. \textit{HPSU} addresses this deficiency.

\section{Conclusion}

In this paper, we introduced a large-scale benchmark called \textit{HPSU}. It was designed to assess human-level perceptual and deep understanding capabilities of \textit{Speech LLMs} in real-world applications. It contained more than 20,000 expert-validated samples in Chinese and English. We utilized a semi-automatic annotation pipeline and yielded the \textit{HPSC} dataset, which contained 50,000 instances. Leveraging this pipeline, \textit{HPSU} probed complex abilities beyond simple tasks. Our comprehensive evaluation of 13 leading models revealed a substantial performance gap between state-of-the-art systems and human proficiency. That indicated current models still struggled to bridge the gap between pattern recognition and genuine reasoning. Our analysis identified current limitations in deep inferential reasoning, robustness against biases, and nuanced cross-lingual comprehension. These findings shown the challenges still existed in the speech task. We believed \textit{HPSU} can help to catalyze research towards truly human-like auditory intelligence.

\section*{Acknowledgments}
This work is supported by the Key-Area Research and Development Program of Guangdong Province (2024B0101050005), National Natural Science Foundation of China (62276279, U22B2060), Guangdong Basic and Applied Basic Research Foundation (2024B1515020032, 2024A1515010253), and Research Foundation of Science and Technology Plan Project of Guangzhou City (2023B01J0001, 2024B01W0004).

\bibliography{main}

\clearpage
\appendix

\subsection{A. Details of original datasets and preprocessing}
\label{appendixA}

\subsubsection{Original Datasets}

\textbf{CelebV-HQ} The video clips in the \textit{CelebV-HQ} dataset are sourced from celebrity videos on online platforms such as YouTube, comprising over 35,000 clips featuring more than 15,000 unique identities. The dataset provides video data with accompanying audio.

\textbf{MAFW} The video clips in the \textit{MAFW} dataset come from a wide range of sources, including film and television content from various countries and regions. It contains over 10,000 video segments with emotional expressions, more than 8,000 of which are in English. The dataset provides video data with accompanying audio.

\textbf{MELD} It contains 1,433 dialogues from the American TV series \textit{Friends}, comprising over 13,000 utterances. Each utterance is accompanied by textual, audio, and visual information. We use a portion of the \textit{MELD} test set as one of the components of \textit{HPSU}, and a portion of the \textit{MELD} train set as one of the components of \textit{HPSC}.

\textbf{VCTK} To collect data from native English speakers with diverse accents, we used the \textit{VCTK} dataset. \textit{VCTK} is a widely used multi-speaker corpus in the speech community, comprising approximately 44,000 speech clips recorded by 110 native English speakers with various accents.

\textbf{MER} We used over 16,000 samples from the \textit{MER2025} dataset as the source of Chinese data. This dataset primarily consists of content from films and television dramas, and includes both visual and audio modalities.

\textbf{The Chinese accent dataset} We supplemented the missing dialectal components in the Chinese data using an internal dataset. This dataset includes audio recordings from speakers of nine major Chinese dialects, such as Mandarin, Northeastern Mandarin, and Southwestern Mandarin.

\subsubsection{Data preprocessing}

In cases where the original dataset did not provide audio files, we first utilized the ffmpeg tool to extract audio in 16 kHz WAV format from the video files. To ensure the quality of the speech data, we employed the speechscore module provided by ClearerVoice-Studio, and evaluated the quality of each extracted audio segment according to the DNSMOS metric \cite{reddy2021dnsmos}. Specifically, each audio segment was scored in four aspects: (i) \textit{Speech Quality (SIG)}; (ii) \textit{Background Noise Quality (BAK)}; (iii) \textit{Overall Quality (OVRL)}; and (iv) \textit{P808 Mean Opinion Score (P808\_MOS)}. It is noteworthy that \textit{DNSMOS} is a non-intrusive assessment method that does not require clean reference audio for evaluation.

As \textit{HPSU} and \textit{HPSC} are centered on speech, we primarily focused on the \textit{SIG} score. To preserve the authenticity of the data as much as possible while ensuring speech quality, we set the \textit{SIG} threshold to 3.5 and required the \textit{BAK} score to be greater than 2.0, ensuring that the background noise remained within an acceptable range. Audio segments that did not meet these thresholds were not immediately discarded; instead, they underwent further processing using the clearvoice module in \textit{ClearerVoice-Studio}. This module leverages the \textit{AV\_MossFormer2\_TSE\_16K} audio-visual fusion model, taking the video as input and utilizing audio-visual information to perform speech enhancement, effectively removing noise unrelated to the target speaker while maximally preserving the original speech characteristics and contextual information of the data. This approach guarantees the quality of the speech data while maintaining its naturalness and authenticity.

Subsequently, all processed data were transcribed using open-source automatic speech recognition models: \textit{Whisper-large-v3} for English and \textit{FireRedASR-AED} for Chinese, resulting in a high-quality tri-modal dataset comprising video, audio, and text. Furthermore, during the subsequent \textit{Experts Check} phase, all video and audio data were rigorously reviewed by experts to ensure that only data meeting the required quality standards were included in the final \textit{HPSU} benchmark.

\subsection{B. SFT on HPSC}
\label{sec:sft}

\begin{table*}[htbp]
\small
\centering
\setlength{\tabcolsep}{3pt}
\renewcommand{\arraystretch}{1.2}
\begin{tabular}{@{} l c *{5}{c} c @{}}
\toprule
\textbf{Lang} & \textbf{Models} 
& \textbf{Social Attributes} 
& \textbf{Emotion Recognition} 
& \textbf{Emotion Reasoning} 
& \textbf{Nonverbal Behavior} 
& \textbf{Discourse Context} 
& \textbf{Avg} \\
\cmidrule(lr){3-7}
& & Age \textbar{} Gen \textbar{} Acc
& MEmo \textbar{} Emos \textbar{} EIns
& SE \textbar{} SH \textbar{} ME \textbar{} MH
& Style \textbar{} Vis \textbar{} Desc
& Intent \textbar{} Scn \textbar{} Subt 
& \\
\midrule

\multirow{3}{*}{EN} 
& Human
& 73.3 \textbar{} 99.8 \textbar{} 67.9
& 85.4 \textbar{} 81.9 \textbar{} 89.1
& 81.9 \textbar{} 78.6 \textbar{} 79.0 \textbar{} 85.0
& 84.2 \textbar{} 90.3 \textbar{} 94.9
& 86.1 \textbar{} 85.3 \textbar{} 89.4
& 84.5 \\
& Qwen-2.5-Omni
& 54.7 \textbar{} 94.6 \textbar{} 23.6
& 49.4 \textbar{} 56.8 \textbar{} 52.9
& 60.9 \textbar{} 47.9 \textbar{} 55.0 \textbar{} 51.0
& 54.5 \textbar{} 57.4 \textbar{} 57.5
& 69.3 \textbar{} 49.4 \textbar{} 72.5
& 56.7 \\
& SFT on HPSC
& \textbf{60.1} \textbar{} \textbf{99.0} \textbar{} \textbf{41.0}
& \textbf{63.0} \textbar{} \textbf{69.4} \textbar{} \textbf{63.3}
& \textbf{68.0} \textbar{} \textbf{61.7} \textbar{} \textbf{67.3} \textbar{} \textbf{61.0}
& \textbf{64.0} \textbar{} \textbf{77.4} \textbar{} \textbf{71.4}
& \textbf{82.0} \textbar{} \textbf{61.3} \textbar{} \textbf{77.2}
& \textbf{67.6} \\
\midrule

\multirow{3}{*}{ZH} 
& Human
& 86.9 \textbar{} 98.2 \textbar{} 70.2
& 95.5 \textbar{} 84.5 \textbar{} 95.4 
& 82.1 \textbar{} 85.6 \textbar{} 89.6 \textbar{} 88.4
& 91.9 \textbar{} 95.9 \textbar{} 97.4
& 91.7 \textbar{} 93.2 \textbar{} 95.9
& 90.2 \\
& Qwen-2.5-Omni
& 53.3 \textbar{} 97.9 \textbar{} 38.0
& 60.1 \textbar{} 40.0 \textbar{} 57.6
& 60.1 \textbar{} 76.3 \textbar{} 45.7 \textbar{} 58.2
& 72.7 \textbar{} 85.0 \textbar{} 79.2
& 78.1 \textbar{} 59.9 \textbar{} 49.2
& 63.2 \\
& SFT on HPSC
& \textbf{61.0 }\textbar{} \textbf{98.1} \textbar{} \textbf{55.0}
& \textbf{79.6} \textbar{} \textbf{66.6} \textbar{} \textbf{66.0}
& \textbf{76.3} \textbar{} \textbf{78.4} \textbar{} \textbf{67.8} \textbar{} \textbf{66.5}
& \textbf{79.7} \textbar{} \textbf{88.8} \textbar{} \textbf{87.1}
& \textbf{82.7} \textbar{} \textbf{66.9} \textbar{} \textbf{52.3}
& \textbf{74.4} \\
\bottomrule
\end{tabular}
\caption{HPSC sft results.}
\label{tab:sft_results}
\end{table*}
\begin{table}[htbp]
    \centering
    \begin{tabular}{cc}
    \toprule
      Models & Speech (\%)\\
      \midrule
      Qwen-2.5-Omni   & 59.86 \\
      SFT on HPSC   & 64.28 \\
      \bottomrule
    \end{tabular}
    \caption{Results on MMAR}
    \label{tab:mmar}
\end{table}
To further enhance the auditory capabilities of the model and validate the effectiveness of our proposed dataset, we conducted a \textit{Supervised Fine-tuning (SFT)} experiment.

\subsubsection{Experimental Setup}
We selected \textit{Qwen2.5-omni-7B}, the best performing open-source model on the \textit{HPSU} benchmark, as our base model. For fine-tuning, we utilized our newly constructed and open-sourced \textbf{HPSC}. Note that the data sources for \textit{HPSC} and \textit{HPSU} are independent, ensuring that there is no overlap between our fine-tuning corpus and the evaluation benchmark.

The fine-tuning was performed on 8 NVIDIA A100 40GB GPUs. During this process, we used an effective batch size of 64, the AdamW optimizer with a learning rate of $10^{-4}$, and trained the model for 2 epochs on the complete dataset.

\subsubsection{Results and Analysis}
As shown in Table~\ref{tab:sft_results}, the performance of the model in all capabilities improved significantly after \textit{SFT}. Specifically, the average accuracy on Chinese and English tasks increased from \textbf{56.7\%} and \textbf{63.2\%} to \textbf{67.9\%} and \textbf{74.4\%}, respectively. Note that with the exception of the accent recognition task, the average scores on all other tasks exceeded the 60\% passing threshold post-SFT. 

Furthermore, to eliminate potential influence from cosourced data and rigorously test the transferability of the model, we also evaluated our HPSC-fine-tuned model on the \textit{MMAR} benchmark. For this test, we used the 294 samples from MMAR's'speech' category. The model accuracy rose from a baseline of 59.86\% to 64.28\% after fine-tuning on \textit{HPSC}. This performance gain on an unseen, out-of-distribution benchmark strongly validates the generalization and robustness of our \textit{HPSC} dataset.

Collectively, these results demonstrate the substantial value of \textit{HPSC}; a dataset with fine-grained annotations that align with human perceptual and cognitive abilities can effectively guide a model to learn more sophisticated and generalizable auditory representations, leading to robust performance gains across diverse benchmarks.

\subsection{C. Details of Evaluated Models and Baselines}
\label{appendixC}
\begin{table*}[!ht]
\small
\renewcommand{\arraystretch}{1.4}
\setlength{\tabcolsep}{1pt}
\begin{tabular}{
    m{82pt} 
    >{\centering\arraybackslash}m{45pt} 
    m{350pt} 
}
\toprule
\textbf{Models} & \textbf{Params} & \textbf{Details} \\
\hline
   Audio Flamingo 2 & 3B & Support Modalities: Inputs$<$Voice, Text$>$, Outputs$<$Text$>$ \par Trained on AudioSkills (4.2M AQA pairs) and LongAudio (260K long audio) datasets. \par Advanced audio understanding and reasoning, supporting 5-minute audio inputs. \\
\hline
   Audio Flamingo 3 & 7B & Support Modalities: Inputs$<$Voice, Text$>$, Outputs$<$Text$>$ \par Trained on AudioSkills-XL (8.7M AQA pairs) and LongAudio-XL (1.26M long audio) datasets. \par Key improvements over AF2 include multi-turn interaction, Chain-of-Thought (CoT) outputs, and extended 10-minute audio support. \\
\hline
   Kimi-Audio-Instruct & 9.7B & Support Modalities: Inputs$<$Voice, Text$>$, Outputs$<$Voice, Text$>$ \par Trained on over 13M hours of audio/text data and fine-tuned on 0.3M hours of task-specific data. \par Supporting audio understanding, generation, and conversation. Analyzing speech as both continuous vectors and discrete tokens. \\
\hline
   Qwen-Audio-Chat & 8B & Support Modalities: Inputs$<$Voice, Text$>$, Outputs$<$Text$>$ \par Fine-tuned on Qwen-Audio with 20K samples. \par A well-designed multi-task training framework enables knowledge sharing across tasks for superior audio understanding. \\
\hline
   Qwen2-Audio-Instruct & 8.4B & Support Modalities: Inputs$<$Voice, Text$>$, Outputs$<$Voice, Text$>$ \par Fine-tuned on Qwen-Audio with 500K hours audio. \par Improving Instruction Following in Qwen-Audio-2 with SFT and DPO. \\
\hline
   SALMONN & 13B & Support Modalities: Inputs$<$Voice, Text$>$, Outputs$<$Text$>$ \par Trained on 44K hours of data covering 12 tasks, including ASR, audio/music QA, and captioning. \par Exhibits cross-modal emergent abilities. Enhanced Semantic and Acoustic Capture via Whisper and Beat Encoders. \\
\hline
   Soundwave & 9B & Support Modalities: Inputs$<$Voice, Text$>$, Outputs$<$Text$>$ \par Fine-tuned on Llama-3.1-8B-Instruct with 10K hours of audio data. \par Superior Speech Understanding from Just 10K Hours of Data via an Efficient Framework. \\
\hline
   Baichuan-Omni-1.5 & 7B & Support Modalities: Inputs$<$Voice, Image, Text$>$, Outputs$<$Voice, Text$>$ \par Trained on 500B tokens of high-quality multimodal data (text, audio, vision). \par An omni-modal model that not only has omni-modal understanding capabilities but also provides end-to-end audio generation capabilities. \\
\hline
   Qwen-2.5-Omni & 7B & Support Modalities: Inputs$<$Voice, Image, Text$>$, Outputs$<$Voice, Text$>$ \par Trained on 800B image/video tokens, 300B audio tokens, and 100B video-with-audio tokens. \par An end-to-end multimodal model designed to perceive diverse modalities while simultaneously generating text and natural speech responses in a streaming manner. \\
\hline
   Gemini 2.5 Flash & --- & Support Modalities: Inputs$<$Voice, Image, Text$>$, Outputs$<$Voice, Text$>$ \par Training data size undisclosed. \par A full-modality Mixture-of-Experts (MoE) model that supports a 1 million token context length. It delivers exceptional reasoning capabilities with minimal computational and latency requirements. \\
\hline
   Audio Flamingo 3.5 & 7B & Support Modalities: Inputs$<$Voice, Text$>$, Outputs$<$Text$>$ \par Training data and model architecture are identical to Audio Flamingo 3. \par This model is identical to AF3 and utilizes a dedicated inference mode, providing a significant advantage in long-context inference. \\
\hline
   Audio-Reasoner & 8.4B & Support Modalities: Inputs$<$Voice, Text$>$, Outputs$<$Voice, Text$>$ \par Fine-tuned on Qwen2-Audio with 1.2M audio reasoning data samples. \par Achieving Superior Reasoning in Qwen2-Audio via Fine-Tuning on High-Quality Data. \\
\hline
   Gemini 2.5 Pro & --- & Support Modalities: Inputs$<$Voice, Image, Text$>$, Outputs$<$Voice, Text$>$ \par Training data size undisclosed. \par A full-modality Mixture-of-Experts (MoE) model that supports a 1 million token context length and possesses powerful capabilities for solving general-purpose problems. \\
\bottomrule
\end{tabular}
\caption{Details of the evaluated models.}
\label{tab:model_summary_full}
\end{table*}

Tables~\ref{tab:model_summary_full} and~\ref{tab:baseline_summary} detail the evaluated models and baselines, respectively. It reveals a significant trend: the state-of-the-art performance, exhibited by both the proprietary \textit{Gemini 2.5 Pro} and the open-source \textit{Qwen-2.5-Omni}, stems not from deep optimization in a single modality, but from their native omni-modal understanding architectures. These models significantly outperform contemporary, specialized models focused solely on the audio domain, such as Kimi-Audio-Insturct. This result strongly suggests that contextualizing audio within a broader multi-modal framework for joint understanding may be the core pathway and future paradigm for advancing the capabilities of large audio models.

\begin{table*}[!t]
\footnotesize
\renewcommand{\arraystretch}{1.5} 
\setlength{\tabcolsep}{6pt} 
\begin{tabular}{
    m{100pt} 
    m{360pt} 
}
\toprule
\textbf{Models} & \textbf{Details} \\
\midrule
   Human & The Human baseline, representing the empirical performance ceiling, is established based on 10 native speakers per language, each holding at least a university degree, who answer 300 stratified-sampled items. Annotators receive comprehensive instructions prior to testing, are isolated during the evaluation process, and must attentively listen to the audio before choosing the most suitable option. \\
\midrule
   Random Guess & The Random Guess baseline is obtained by randomly selecting an answer option for each item, simulating chance-level performance. \\
\midrule
   Whisper+ChatGPT-4o & The Whisper+ChatGPT-4o baseline is established by first transcribing each audio item using the Whisper-large-v3 model; only the transcription and question, and not the audio, are provided to ChatGPT-4o(2024-11-20) for answer selection. \\
\bottomrule
\end{tabular}
\caption{Details of baselines.}
\label{tab:baseline_summary}
\end{table*}


\subsection{D. Experimental Configuration}

\subsubsection{Evaluated Models of HPSU}



We selected 11 leading open-source systems for the English section: \textit{Audio Flamingo 2 (AF2)}, \textit{Audio Flamingo 3 (AF3)}, \textit{Kimi-Audio-Instruct (Kimi-Audio)}, \textit{Qwen-Audio-Chat (Qwen-Audio)}, \textit{Qwen2-Audio-Instruct (Qwen2-Audio)}, \textit{SALMONN}, \textit{Soundwave}, \textit{Baichuan-Omni-1.5 (BC-Omni-1.5)}, \textit{Qwen2.5-Omni}, \textit{Audio Flamingo 3.5 (AF3.5)}, and \textit{Audio-Reasoner}, as well as 2 proprietary models (\textit{Gemini 2.5 Flash and Pro}).

Since \textit{Audio Flamingo 2}, \textit{Audio Flamingo 3.5}, and \textit{Soundwave} do not provide stable support for Chinese, we did not report the results of these models in the Chinese section of HPSU.

\subsubsection{Tasks Used in the Induction Test}

Due to differences in prompt formats, induction testing was only incorporated into \textit{Single-Choice (SC)} tasks; \textit{Judgment/Binary (JB)} and \textit{Multiple-Choice (MC)} tasks, which consist of \textit{Emotions}, \textit{Description}, and \textit{Subtext}, were not included.
\begin{figure}[!h]
  \centering
  \includegraphics[width=0.8\linewidth]{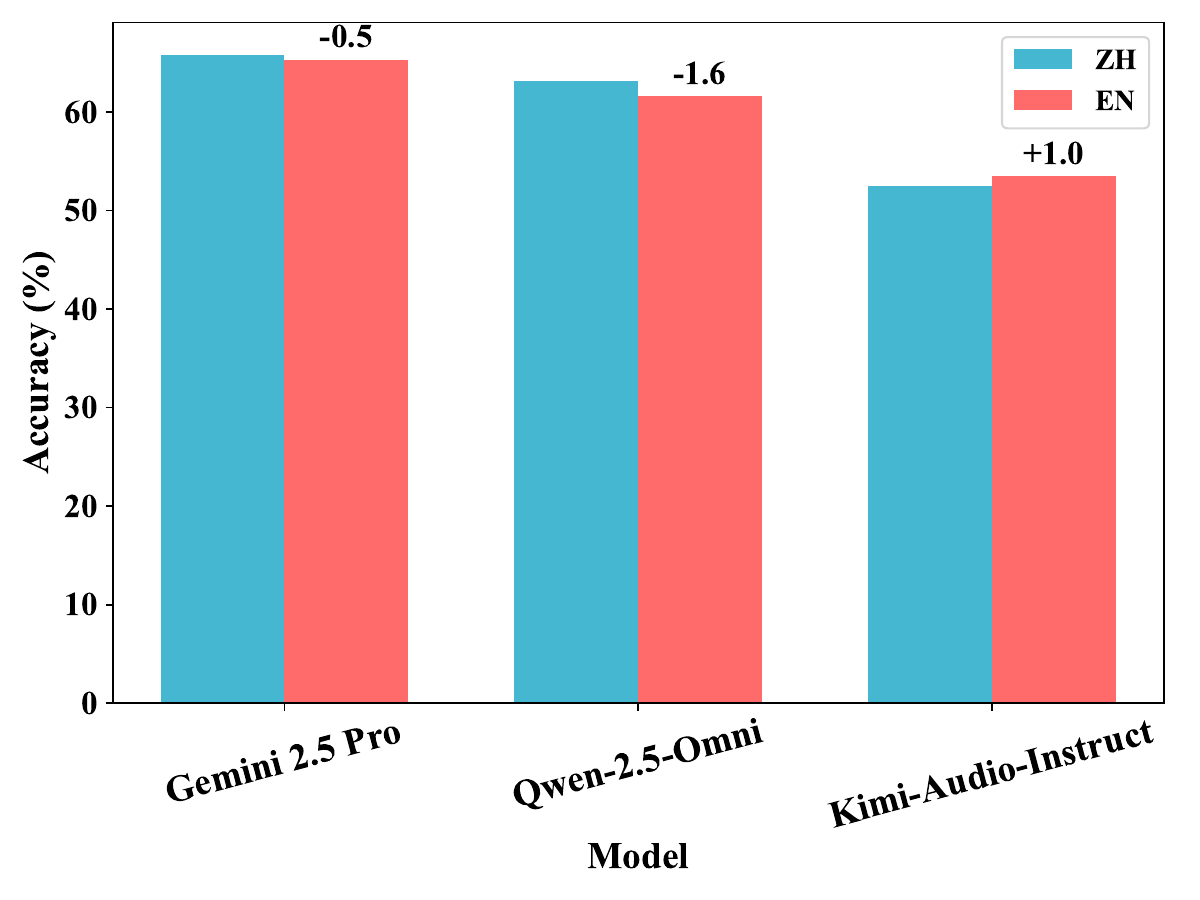}
  \caption{Average accuracy (\%) of different models on the Chinese subset of HPSU under Chinese and English prompts. EN denotes evaluation with English prompts and English reference answers, while ZH denotes evaluation with Chinese prompts and Chinese reference answers. The numeric labels on top of the bars indicate the difference (EN - ZH) for each model.}
  \label{fig:hpsu_chinese_subset_bar}
\end{figure}

\subsection{E. Performance Comparison of Major Models on Cross-Lingual Instruction Following}




To rule out the potential impact of instruction language on model performance, we conducted tests using cross-lingual instructions. Specifically, on the Chinese subset, we compared results obtained with two settings: (i) the audio, prompt, and reference answer are all in Chinese; and (ii) the audio is in Chinese, while the prompt and reference answer are both in English. We selected two high-performing open-source models, \textit{Kimi-Audio-Instruct} and \textit{Qwen-2.5-Omni}, as well as the best-performing proprietary model, \textit{Gemini 2.5 Pro}, and conducted evaluations on the ZH subset of \textit{HPSU}. The results are presented in Figure~\ref{fig:hpsu_chinese_subset_bar}. Our results demonstrate that cross-lingual (Chinese and English) instructions have minimal impact on model evaluation performance, with the average accuracy difference not exceeding 2\%.

\end{document}